\documentclass{article}
\pdfoutput=1

\usepackage[final]{neurips_2024}

\usepackage[utf8]{inputenc} 
\usepackage[T1]{fontenc}    
\usepackage{hyperref}       
\usepackage{url}            
\usepackage{booktabs}       
\usepackage{amsfonts}       
\usepackage{nicefrac}       
\usepackage{microtype}      
\usepackage{xcolor}         
\usepackage{graphicx}
\usepackage{subcaption}
\usepackage{natbib}
\title{Darkit: A User-Friendly Software Toolkit for Spiking Large Language Model}

\author{
  Xin Du \footnotemark[2] \footnotemark[3]  ,
  Shifan Ye \footnotemark[3]  ,
  Qian Zheng\footnotemark[3] \footnotemark[4]  ,
  Yangfan Hu\footnotemark[3] \footnotemark[4]  ,
  Rui Yan \footnotemark[3] \footnotemark[4] ,
  Shunyu Qi \footnotemark[3] \footnotemark[4] , \\
\textbf{  Shuyang Chen \footnotemark[3] \footnotemark[4] , 
  Huajin Tang \footnotemark[3] \footnotemark[4] , 
  Gang Pan\thanks{Corresponding author.} \thanks{School of Software Technology, Zhejiang University, Zhejiang, China.}  \thanks{College of Computer Science and Technology, Zhejiang University, Zhejiang, China.}   \thanks{The State Key Lab of Brain-Machine Intelligence, Zhejiang University, Zhejiang, China.} \ and  Shuiguang Deng \footnotemark[1] 
  \footnotemark[3] \footnotemark[4]
  }
}

\begin{document}

\maketitle



Large language models (LLMs) have been widely applied in various practical applications, typically comprising billions of parameters, with inference processes requiring substantial energy and computational resources~\cite{brown2020language,touvron2023llama}. In contrast, the human brain, employing bio-plausible spiking mechanisms, can accomplish the same tasks while significantly reducing energy consumption, even with a similar number of parameters~\cite{xing2024spikellm}. Based on this, several pioneering researchers have proposed and implemented various large language models that leverage spiking neural networks~\cite{xing2024spikelm,zhu2023spikegpt}. They have demonstrated the feasibility of these models, validated their performance, and open-sourced their frameworks and partial source code. To accelerate the adoption of brain-inspired large language models and facilitate secondary development for researchers, we are releasing a software toolkit named DarwinKit (Darkit). The toolkit is designed specifically for learners, researchers, and developers working on spiking large models, offering a suite of highly user-friendly features that greatly simplify the learning, deployment, and development processes.

When novices attempt to use existing open-source frameworks and tools for the study and development of spiking large language models, they often face several troubles that result in significant time consumption:
\begin{enumerate}
    \item \textbf{Configuration and Preprocessing:} Configuring the development environment, preprocessing data, and preparing a suitable tokenizer according to the requirements of corresponding spiking large language model frameworks can be cumbersome and time-consuming.
    \item \textbf{Parameter Tuning and Testing:} Tuning and testing the parameters of spiking large language models are often complex and tedious. Comparing different models and parameter configurations necessitates frequent code updates and continuous monitoring of training and inference processes.
    \item \textbf{Secondary Development:} Conducting secondary development based on the source code of existing spiking large language models requires an in-depth understanding of the computational graph of the original framework and the relationships between individual blocks. This results in substantial learning and analysis costs when navigating the original codebase.
    \item \textbf{Data Monitoring and Experiment Comparison:} For researchers performing secondary development, the ability to view, record, and trace model runtime data in real-time is essential. Comparing experimental results across different configurations often requires additional tools to visualize and analyze the outcomes.
\end{enumerate}

Darkit introduces the following ten features to address the above issues commonly faced by researchers and developers working with spiking large language models:

\textbf{1. One-Click Environment Configuration:}  Users can configure the development environment with a single command, requiring only CUDA and Conda environments on their local machines. As shown in Figure 1, the toolchain can be installed effortlessly via pip.
\begin{figure}[!ht]
    \centering
    \includegraphics[width=0.85\textwidth]{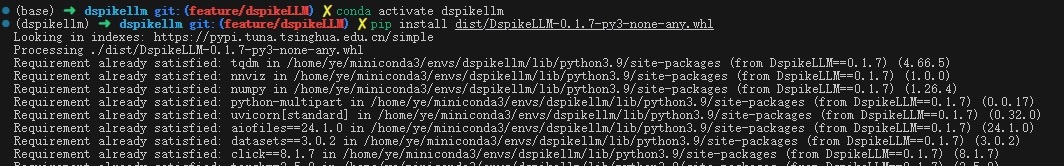} 
    \caption{The toolchain is installed using the pip command.} 
    \label{fig:sample_figure} 
    \vspace{-0.5cm}
\end{figure}

\textbf{2. Integrated Preprocessed Datasets:} Darkit includes preprocessed datasets such as Wikitext, Wikipedia, Ultrachat, and Fineweb, which are readily available via code or a graphical interface on the web (Figure 2). The tool will continue to maintain and expand its dataset collection while providing interfaces for users to contribute new datasets for integration.
\begin{figure}[!ht]
    \centering
    \begin{subfigure}[b]{0.48\textwidth} 
        \centering
        \includegraphics[width=\textwidth, height=2.8cm]{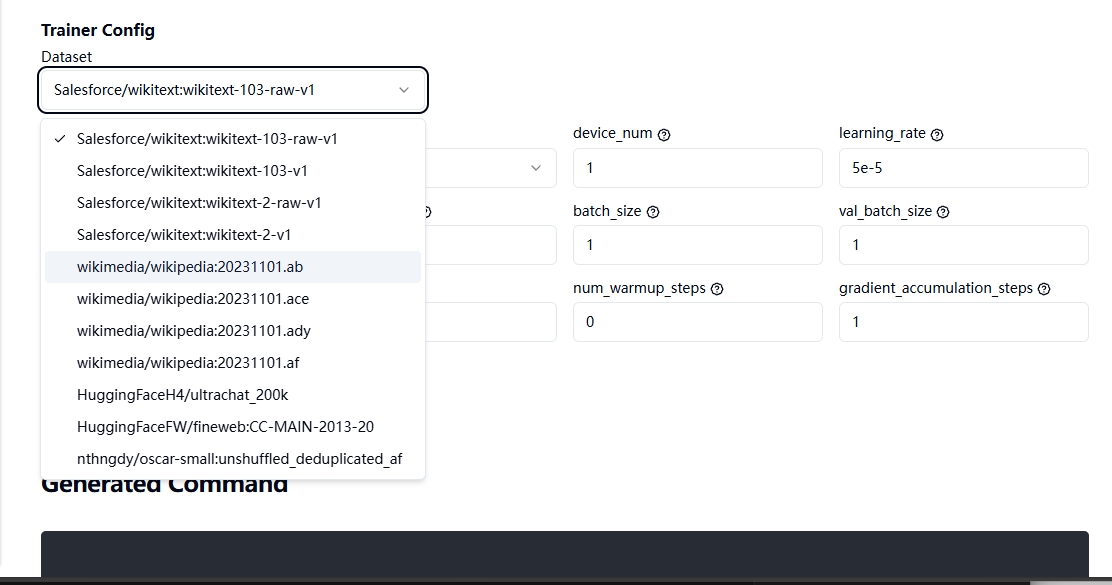} 
        \caption{Datasets displayed via graphical interface} 
        \label{fig:subfigure1} 
    \end{subfigure}
    \hfill 
    \begin{subfigure}[b]{0.48\textwidth}
        \centering
        \includegraphics[width=\textwidth, height=2.5cm]{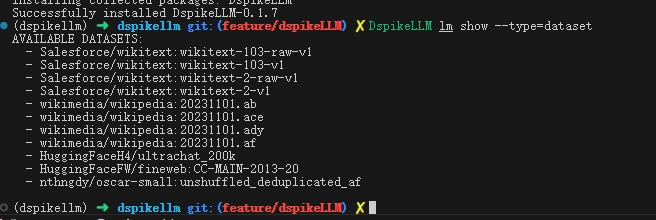} 
        \caption{Datasets displayed in coding terminal}
        \label{fig:subfigure2}
    \end{subfigure}
    
    \caption{The display of integrated preprocessed datasets.} 
    \label{fig:overall_figure} 
\end{figure}

\textbf{3. Integrated Preprocessed Tokenizers:} Popular tokenizers, including GPT-2 (small, medium, and large variants), BERT-based-cased, and BERT-base-Chinese, are pre-integrated in Darkit (as shown in Figure 3). Similar to datasets, tokenizers will be continuously updated and allow users to add new tokenizers through the provided interfaces.
\begin{figure}[!ht]
    \centering
    \begin{subfigure}[b]{0.48\textwidth} 
        \centering
        \includegraphics[width=\textwidth, height=2.8cm]{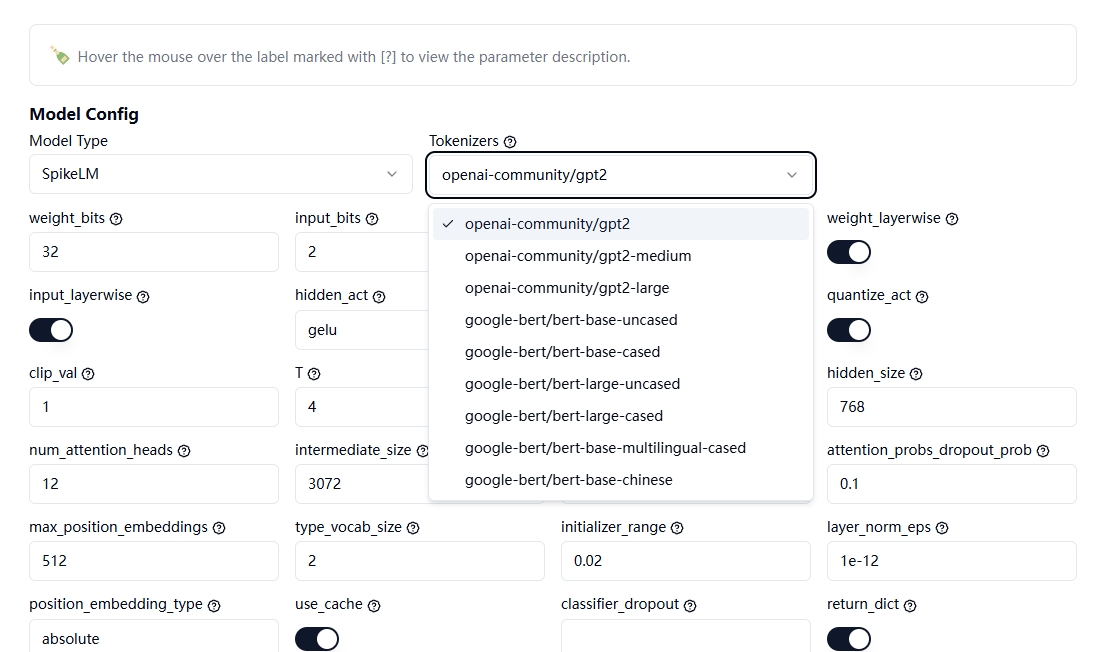} 
        \caption{Tokenizers displayed via graphical interface} 
        \label{fig:subfigure1} 
    \end{subfigure}
    \hfill 
    \begin{subfigure}[b]{0.48\textwidth}
        \centering
        \includegraphics[width=\textwidth, height=2.5cm]{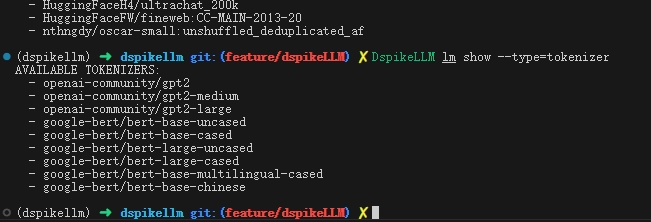} 
        \caption{Tokenizers displayed in coding terminal}
        \label{fig:subfigure2}
    \end{subfigure}
    
    \caption{The display of integrated preprocessed tokenizers.} 
    \label{fig:overall_figure} 
\end{figure}
\begin{figure}[!ht]
    \centering
    \begin{subfigure}[b]{0.46\textwidth} 
        \centering
        \includegraphics[width=\textwidth, height=3.8cm]{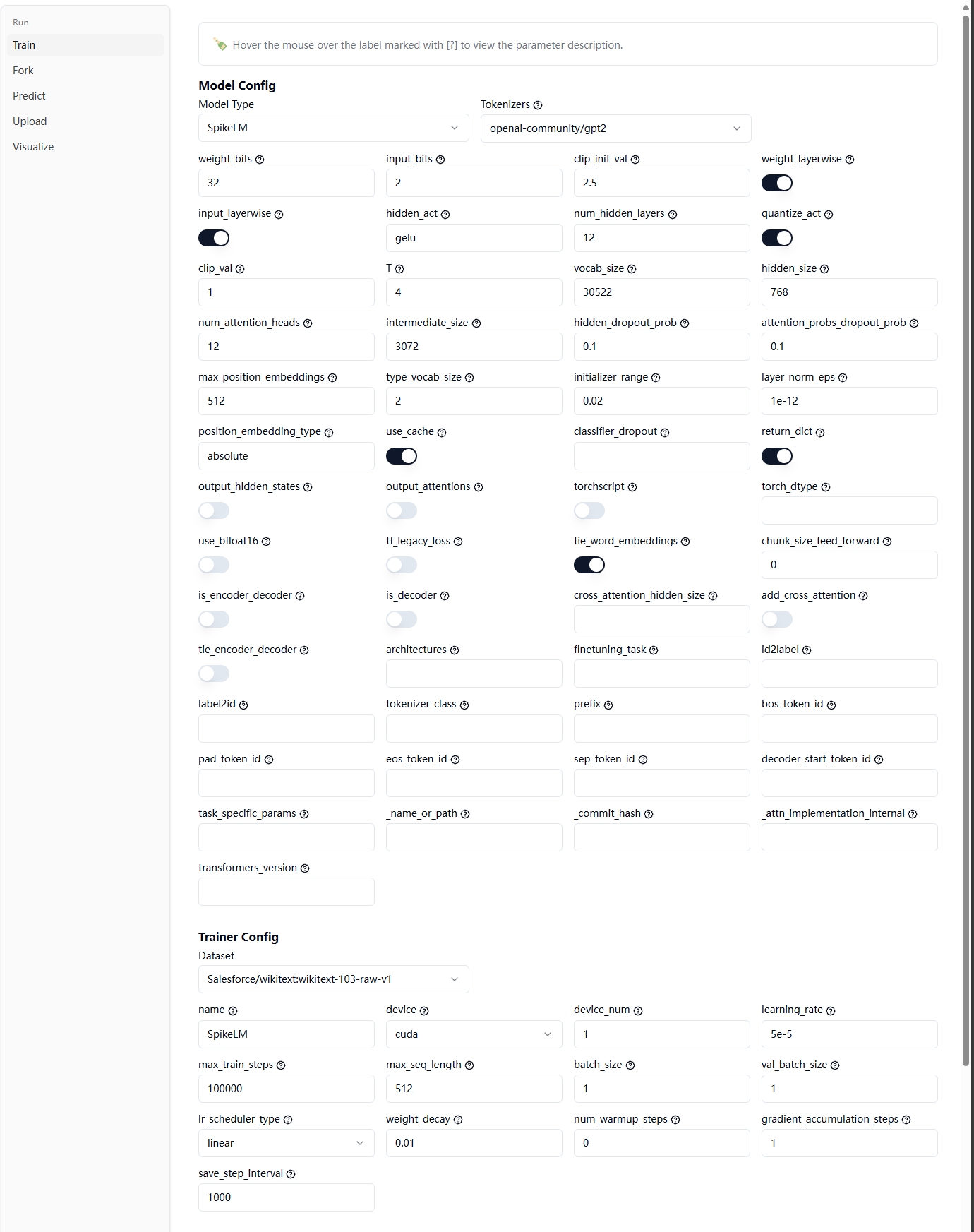} 
        \caption{The process of model training} 
        \label{fig:subfigure1} 
    \end{subfigure}
    \hfill 
    \begin{subfigure}[b]{0.46\textwidth}
        \centering
        \includegraphics[width=\textwidth, height=3.8cm]{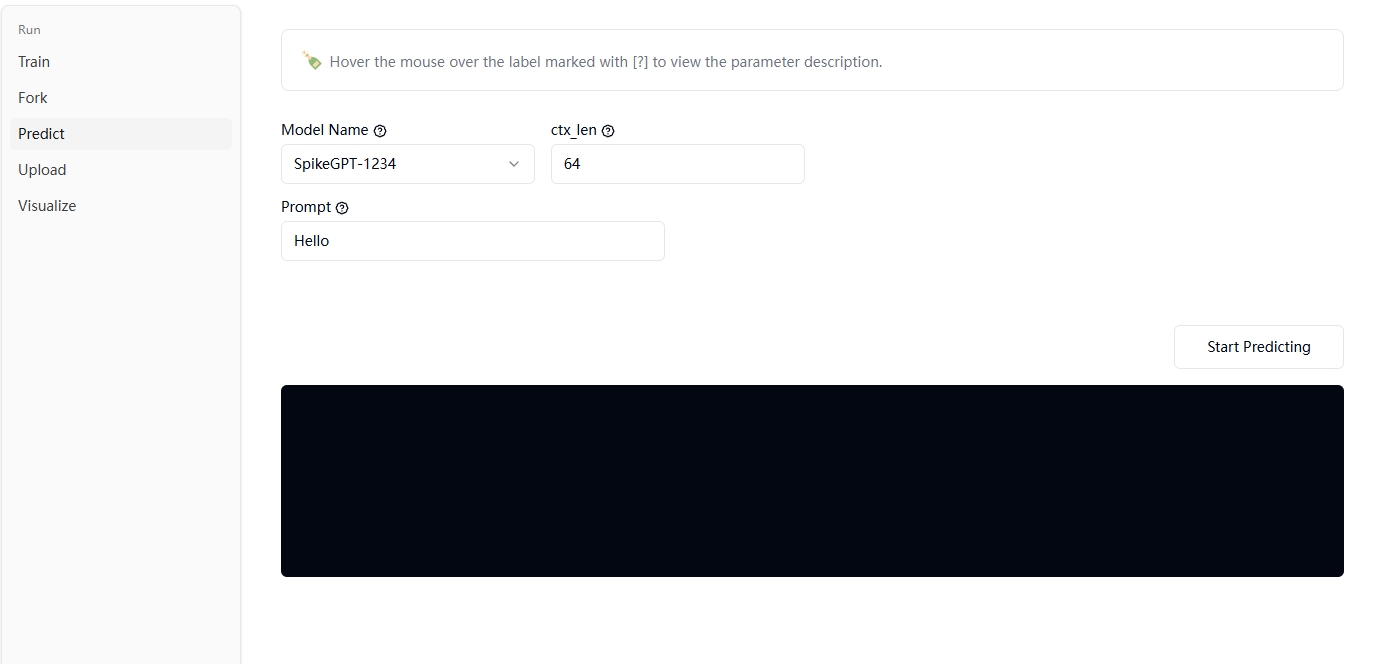} 
        \caption{The process of model inference}
        \label{fig:subfigure2}
    \end{subfigure}
    
    \caption{GUI-based tool to automate the generation of tuning and testing commands.} 
    \label{fig:overall_figure} 
\end{figure}

\textbf{4. GUI-Based Command Generator for Tuning and Testing: } Darkit includes a user-friendly GUI tool to automate the generation of tuning and testing commands (Figure 4). For instance, users can select basic settings (e.g., model name, dataset, and parameters) from dropdown menus. The tool dynamically updates model-specific parameters based on the selection, enabling users to configure the parameter search space easily. Finally, it generates the corresponding tuning command, which can be directly copied and executed in the terminal.

\textbf{5. Real-Time Monitoring and Visualization: } During model training and inference, users can view and record data in real time through both web and code interfaces (Figure 5). Darkit also provides visualization tools for monitoring model performance and understanding data flow.
\begin{figure}[!ht]
    \centering
    \begin{subfigure}[b]{0.48\textwidth} 
        \centering
        \includegraphics[width=\textwidth, height=3.8cm]{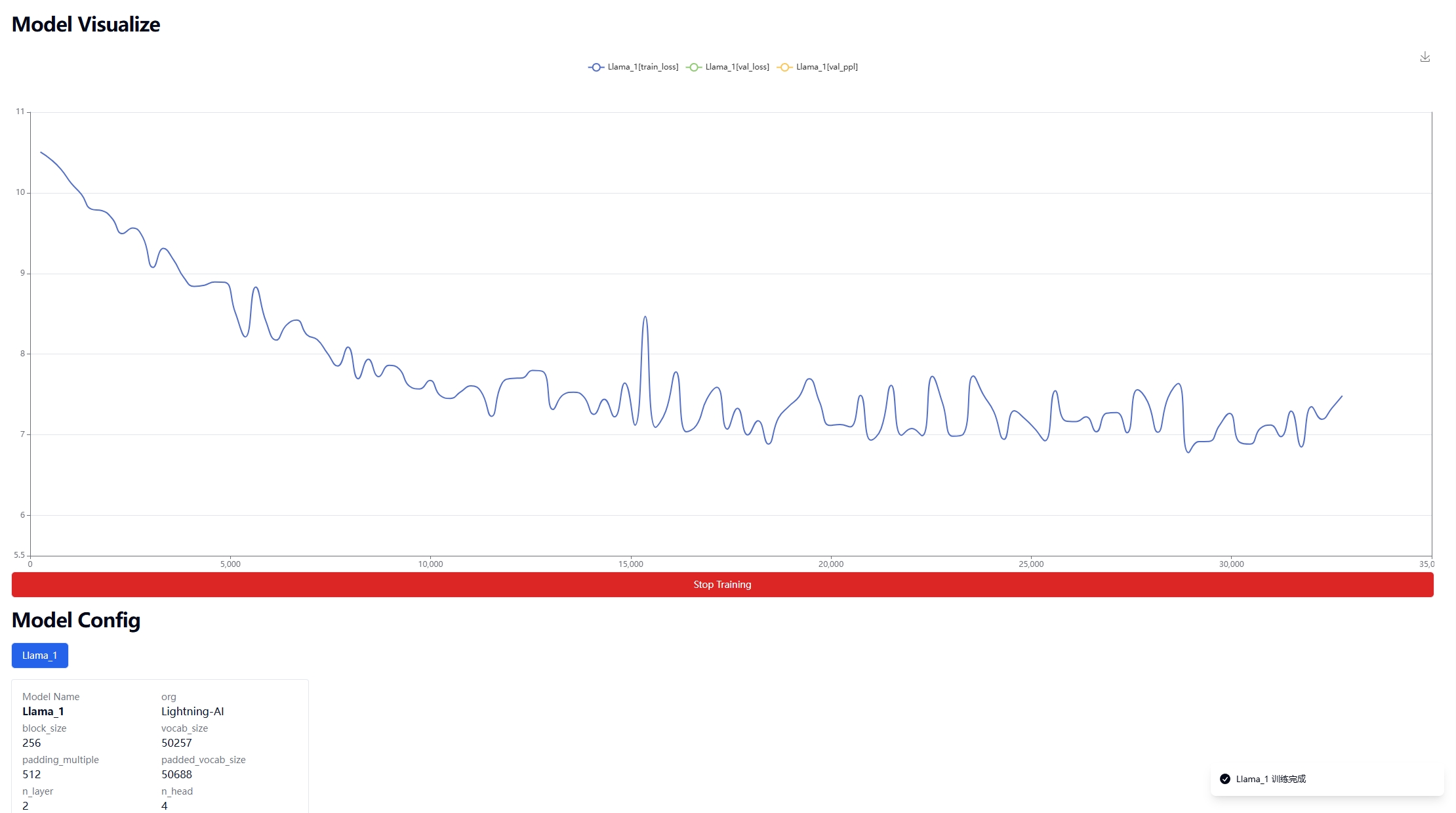} 
        \caption{Recorded data displayed via graphical interface} 
        \label{fig:subfigure1} 
    \end{subfigure}
    \hfill 
    \begin{subfigure}[b]{0.48\textwidth}
        \centering
        \includegraphics[width=\textwidth, height=3.8cm]{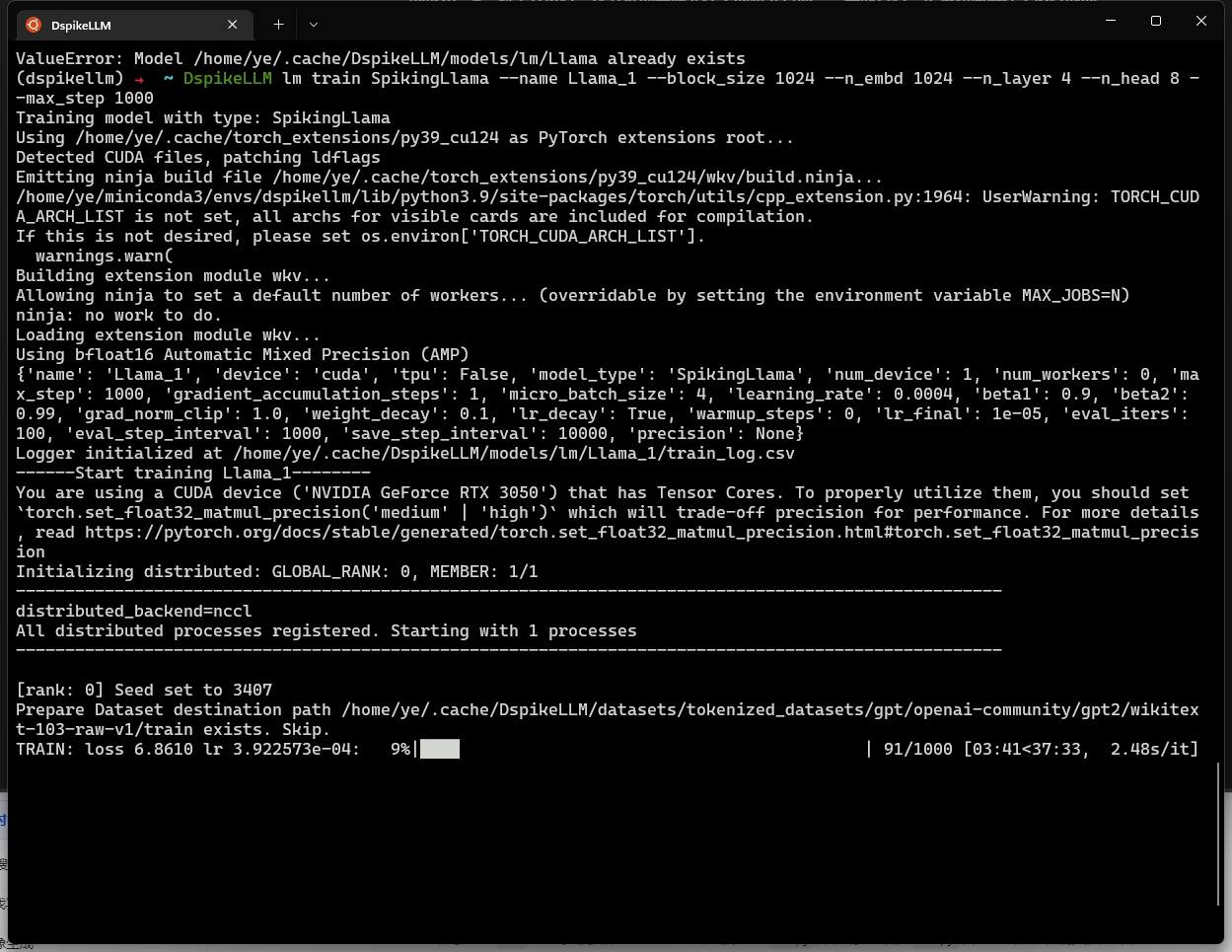} 
        \caption{Recorded data displayed via coding terminal}
        \label{fig:subfigure2}
    \end{subfigure}
    
    \caption{Recorded data displayed in real-time monitoring and visualization.} 
    \label{fig:overall_figure} 
\end{figure}

\textbf{6. Automated Extraction of Model Computational Graphs: } Darkit automatically extracts the computational graph of large models, organizing nodes into a hierarchical tree structure based on module identifiers (Figure 6a). This feature allows users to explore the model architecture interactively, making it easier to understand than reading the raw source code (Figure 6b). When a module is selected on the web frontend, the backend extracts the relevant code segment and displays it for the user, facilitating comprehension of the module's functionality without delving into the source code. 
\begin{figure}[!ht]
    \centering
    \begin{subfigure}[b]{0.48\textwidth} 
        \centering
        \includegraphics[width=\textwidth, height=3.8cm]{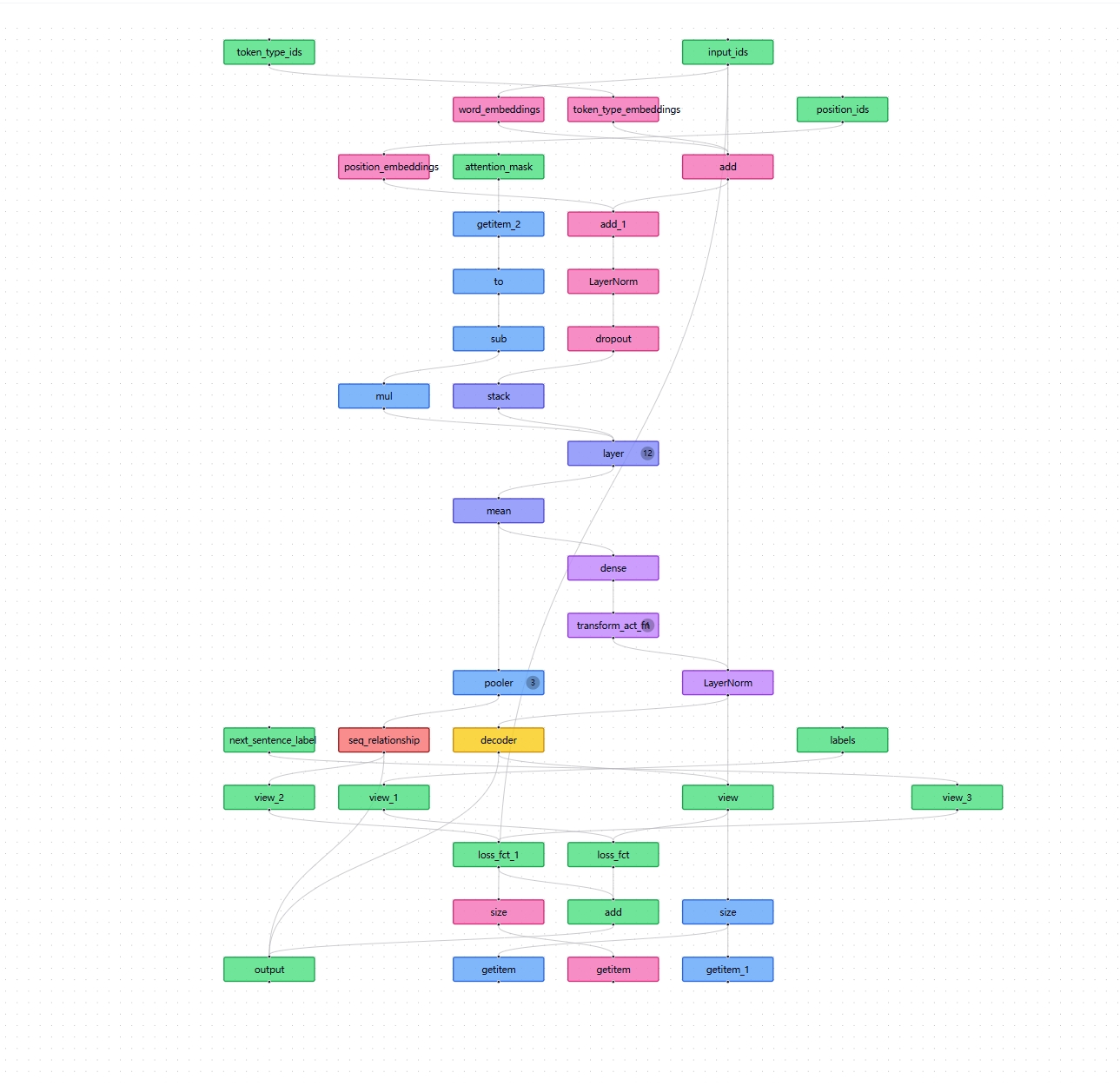} 
        \caption{A hierarchical tree representation of a large model} 
        \label{fig:subfigure1} 
    \end{subfigure}
    \hfill 
    \begin{subfigure}[b]{0.48\textwidth}
        \centering
        \includegraphics[width=\textwidth, height=3.8cm]{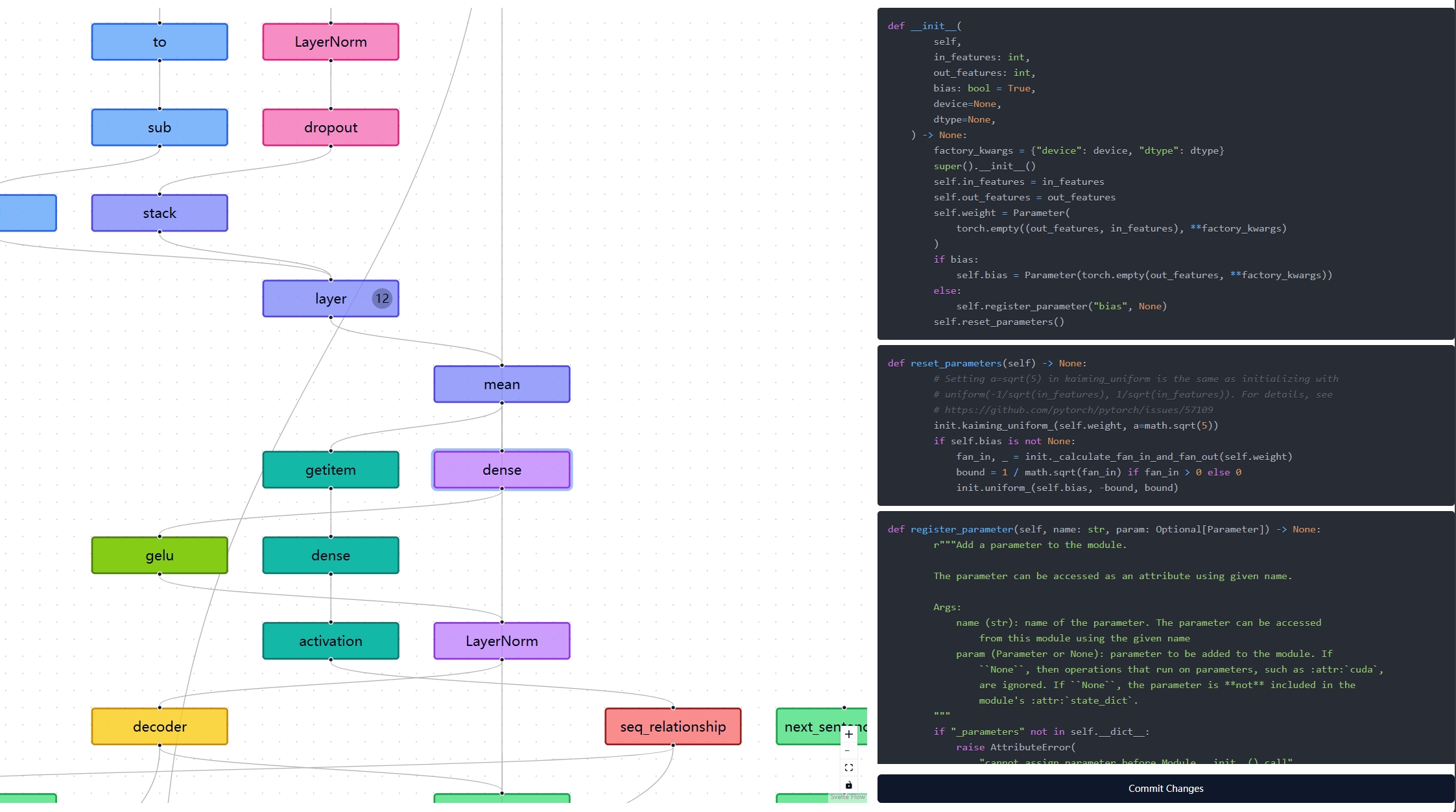} 
        \caption{Accessing raw source code by clicking blocks}
        \label{fig:subfigure2}
    \end{subfigure}
    
    \caption{Automated extraction and display of model computational graphs.} 
    \label{fig:overall_figure} 
\end{figure}
\begin{figure}[!ht]
    \centering
    \begin{subfigure}[b]{0.48\textwidth} 
        \centering
        \includegraphics[width=\textwidth, height=3.8cm]{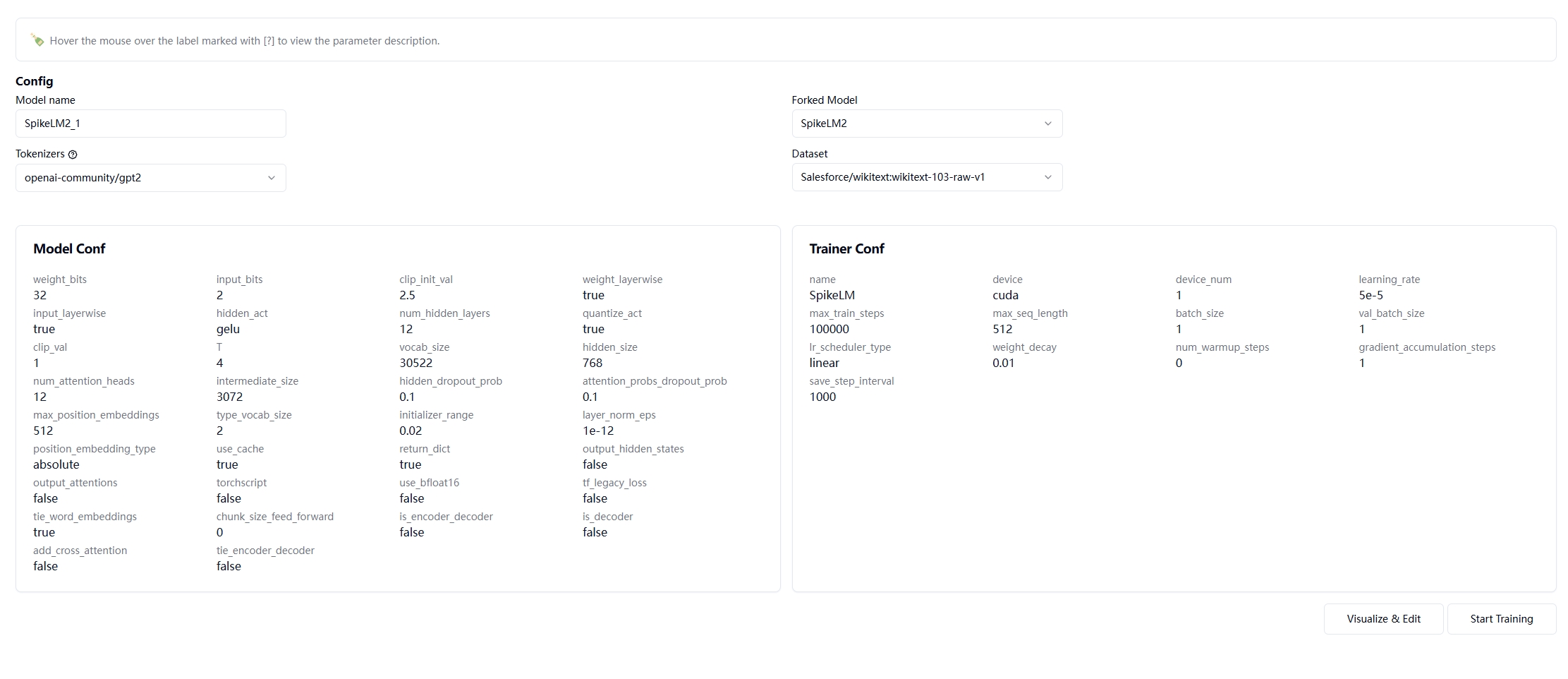} 
        \caption{The code editing and re-injecting} 
        \label{fig:subfigure1} 
    \end{subfigure}
    \hfill 
    \begin{subfigure}[b]{0.48\textwidth}
        \centering
        \includegraphics[width=\textwidth, height=3.8cm]{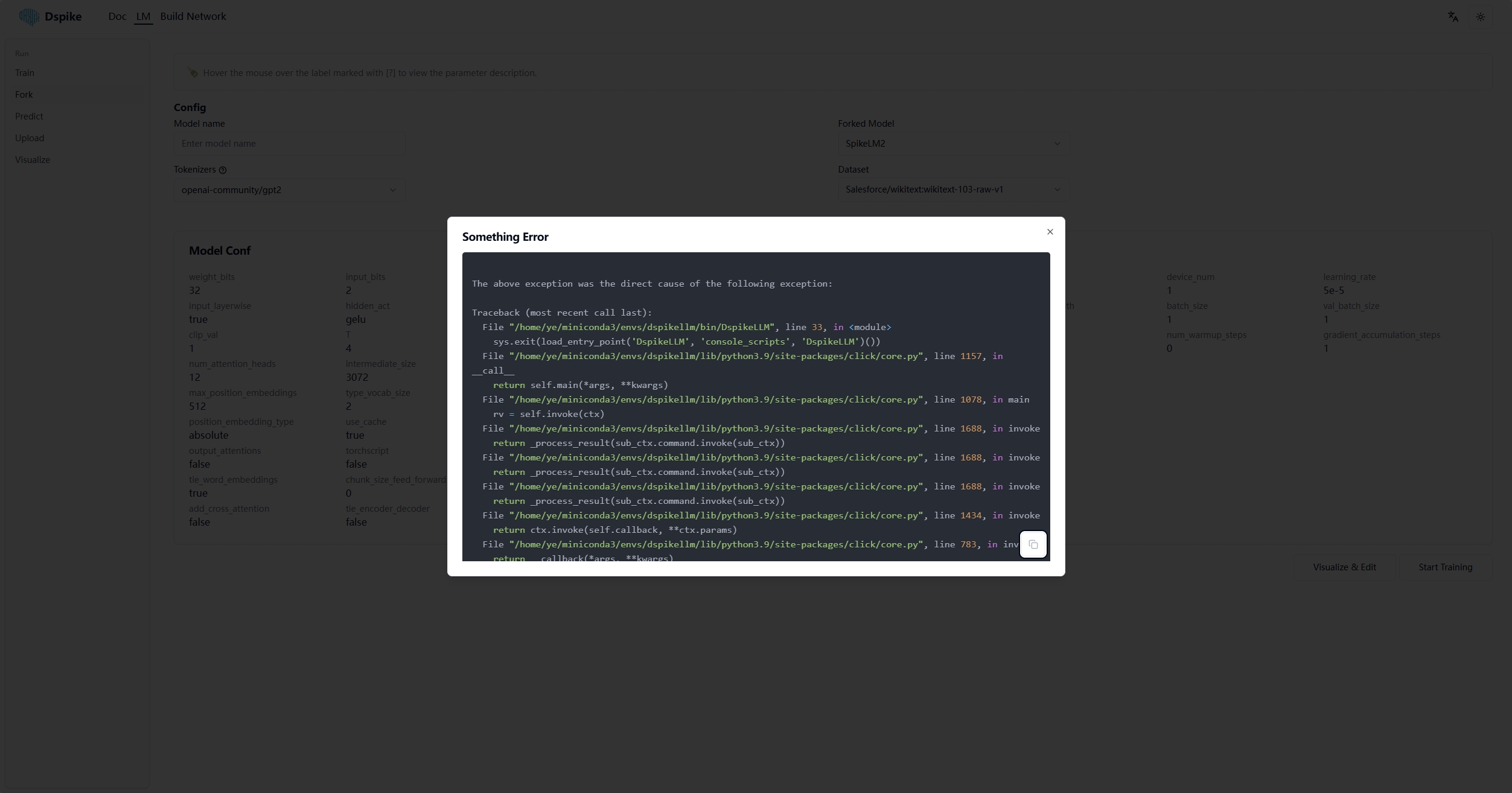} 
        \caption{The validation of user's code editing}
        \label{fig:subfigure2}
    \end{subfigure}
    
    \caption{Code editing, re-integration and validation process.} 
    \label{fig:overall_figure} 
\end{figure}

\textbf{7. Code Editing and Re-Integration: } Darkit provides interfaces for users to edit extracted source code via the web frontend or backend. The modified code can be re-injected into the model for convenient architectural and computational flow adjustments (Figure 7a). The tool validates user edits, offers suggestions for corrections, and supports saving, training, and executing the modified models directly (Figure 7b).

\textbf{8. Flowchart-Based Model Design:  } Darkit allows users to design spiking large language model computational graphs through a visual flowchart interface or terminal commands. The combination of graphical and coding interfaces facilitates the creation and implementation of spiking neural network models (Figure 8).
\begin{figure}[!ht]
    \centering
    \begin{subfigure}[b]{0.48\textwidth} 
        \centering
        \includegraphics[width=\textwidth, height=3.8cm]{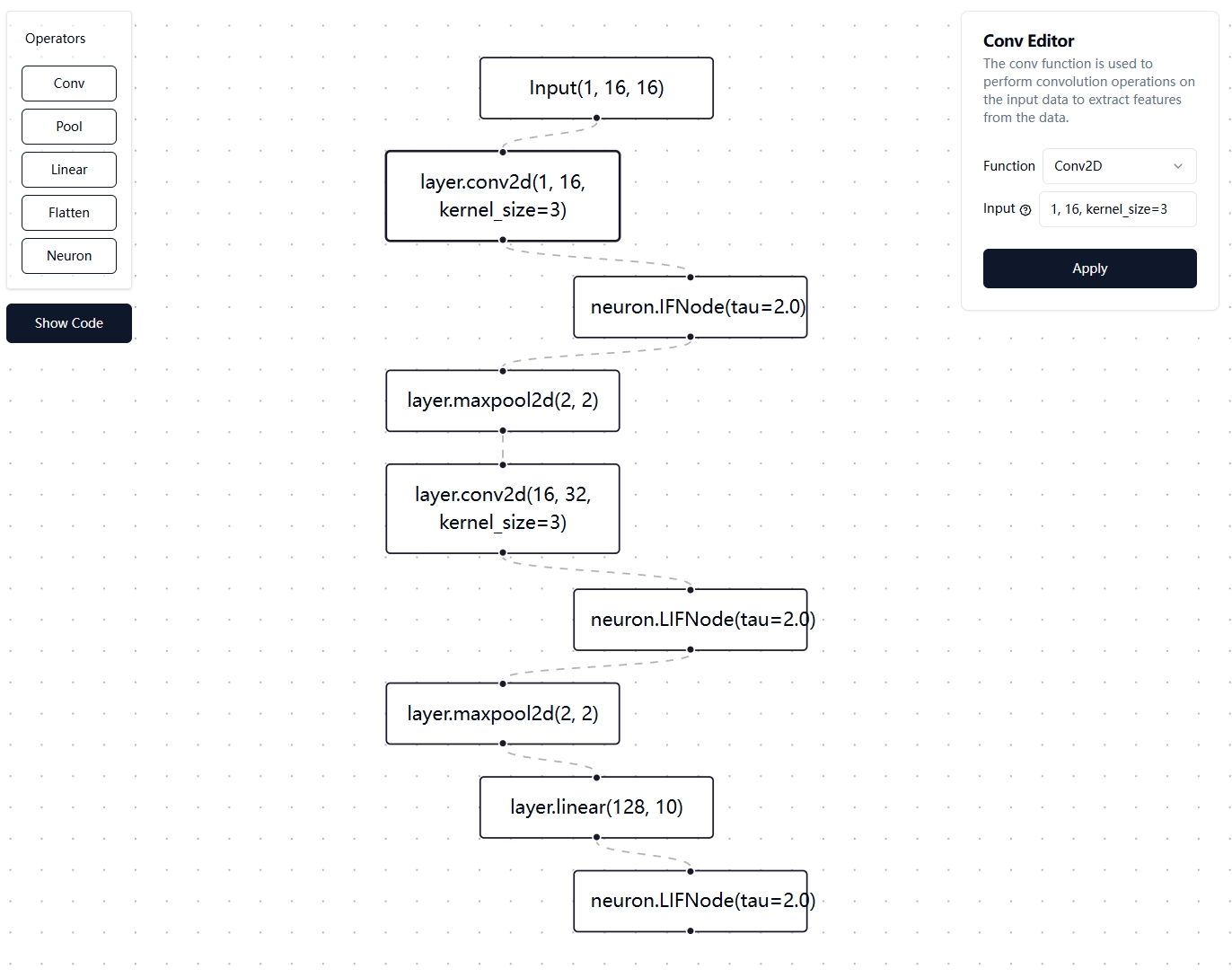} 
        \caption{Designing a model graph with a visual flowchart} 
        \label{fig:subfigure1} 
    \end{subfigure}
    \hfill 
    \begin{subfigure}[b]{0.48\textwidth}
        \centering
        \includegraphics[width=\textwidth, height=3.8cm]{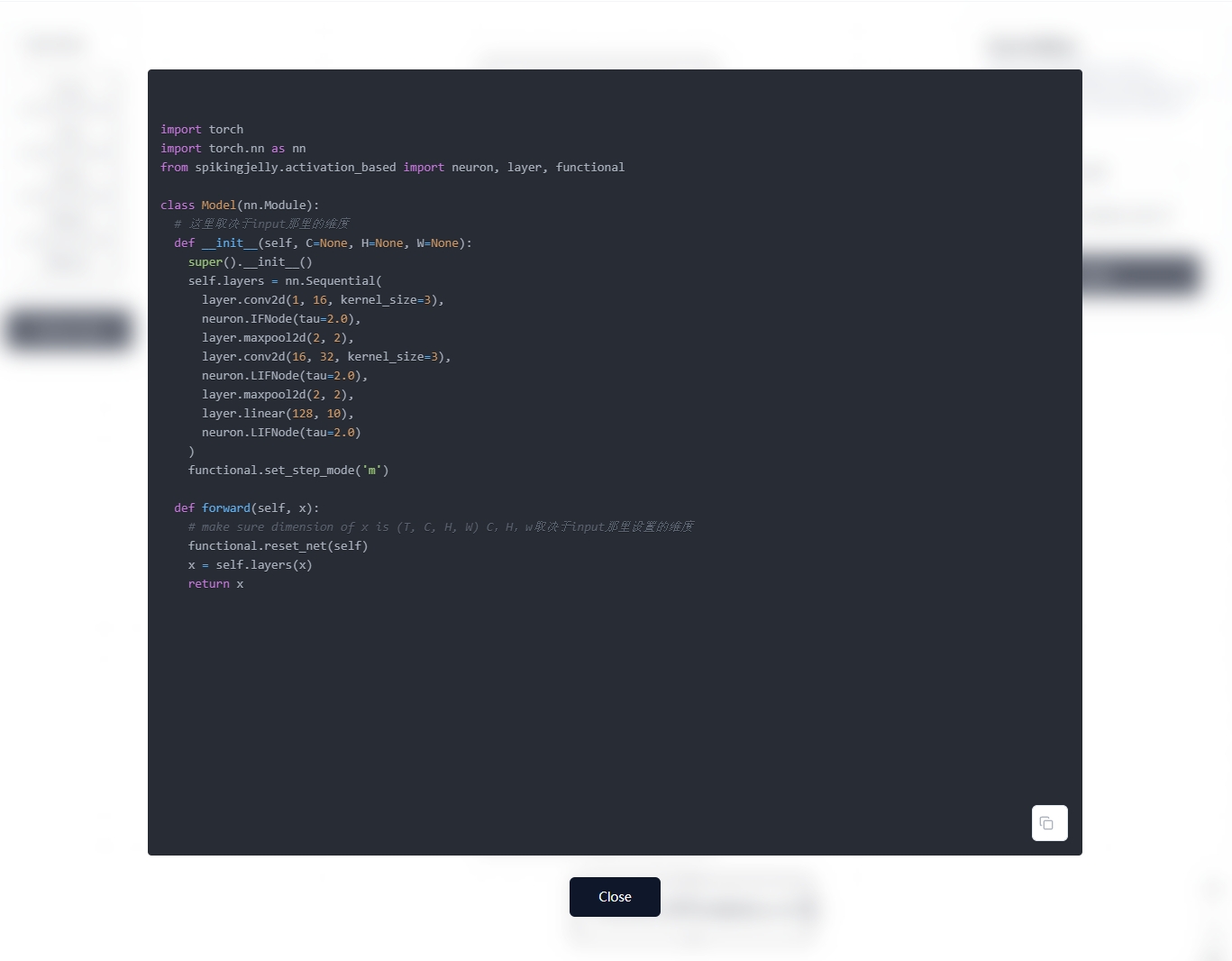} 
        \caption{Integration of graphical and coding interfaces}
        \label{fig:subfigure2}
    \end{subfigure}
    
    \caption{Illustration of flowchart-based model design process.} 
    \label{fig:overall_figure} 
\end{figure}

\textbf{9. Comprehensive Logging and Visualization Tools: } To enhance transparency and traceability, Darkit includes robust logging and visualization capabilities. These tools automatically organize, trace, and analyze issues encountered during training and inference. Additionally, the saved logs can be used to generate comparative experiment charts (Figure 9).
\begin{figure}[!ht]
    \centering
    \begin{subfigure}[b]{0.48\textwidth} 
        \centering
        \includegraphics[width=\textwidth, height=3.8cm]{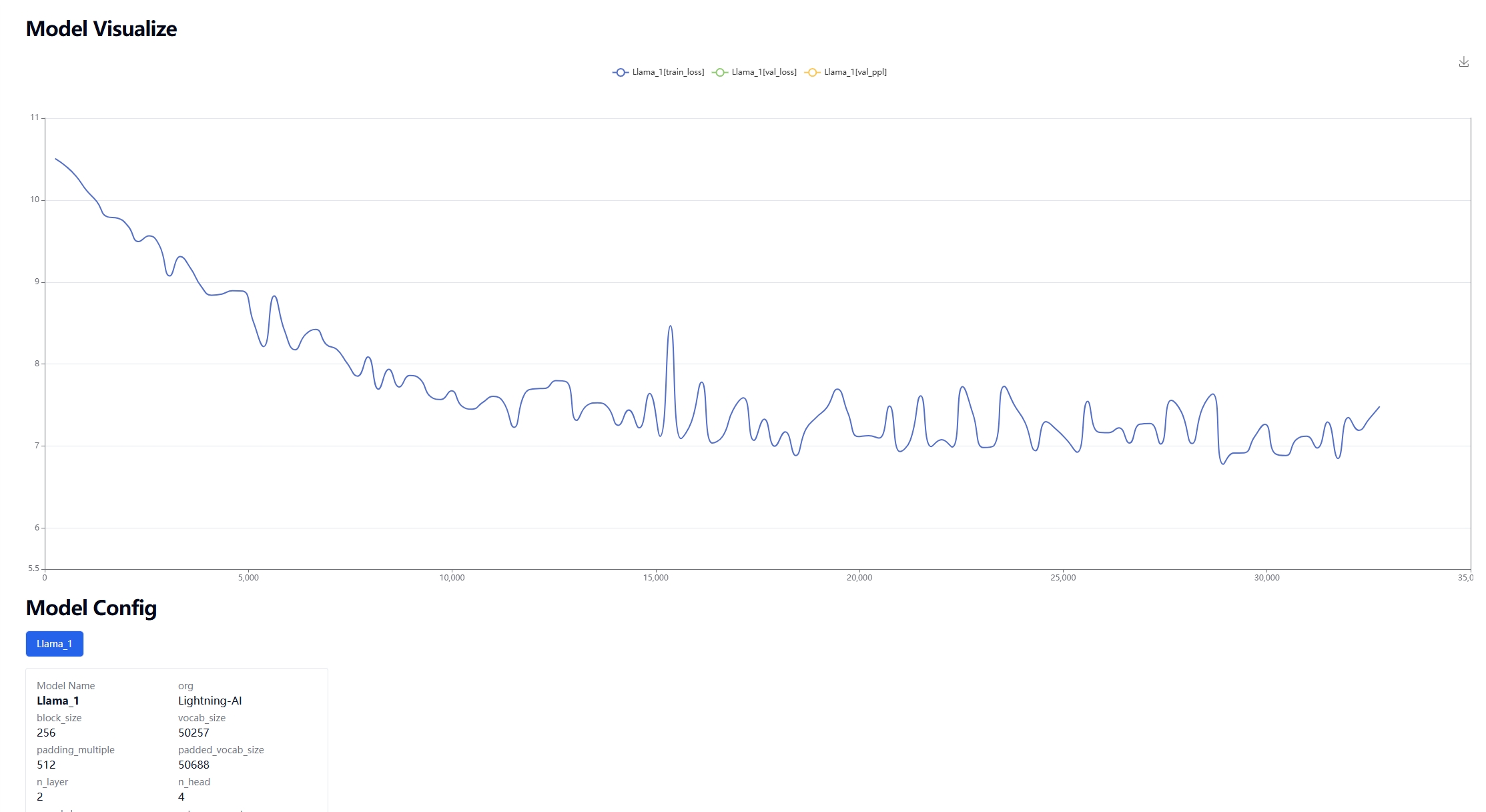} 
        \caption{Visualization of data for a single model} 
        \label{fig:subfigure1} 
    \end{subfigure}
    \hfill 
    \begin{subfigure}[b]{0.48\textwidth}
        \centering
        \includegraphics[width=\textwidth, height=3.8cm]{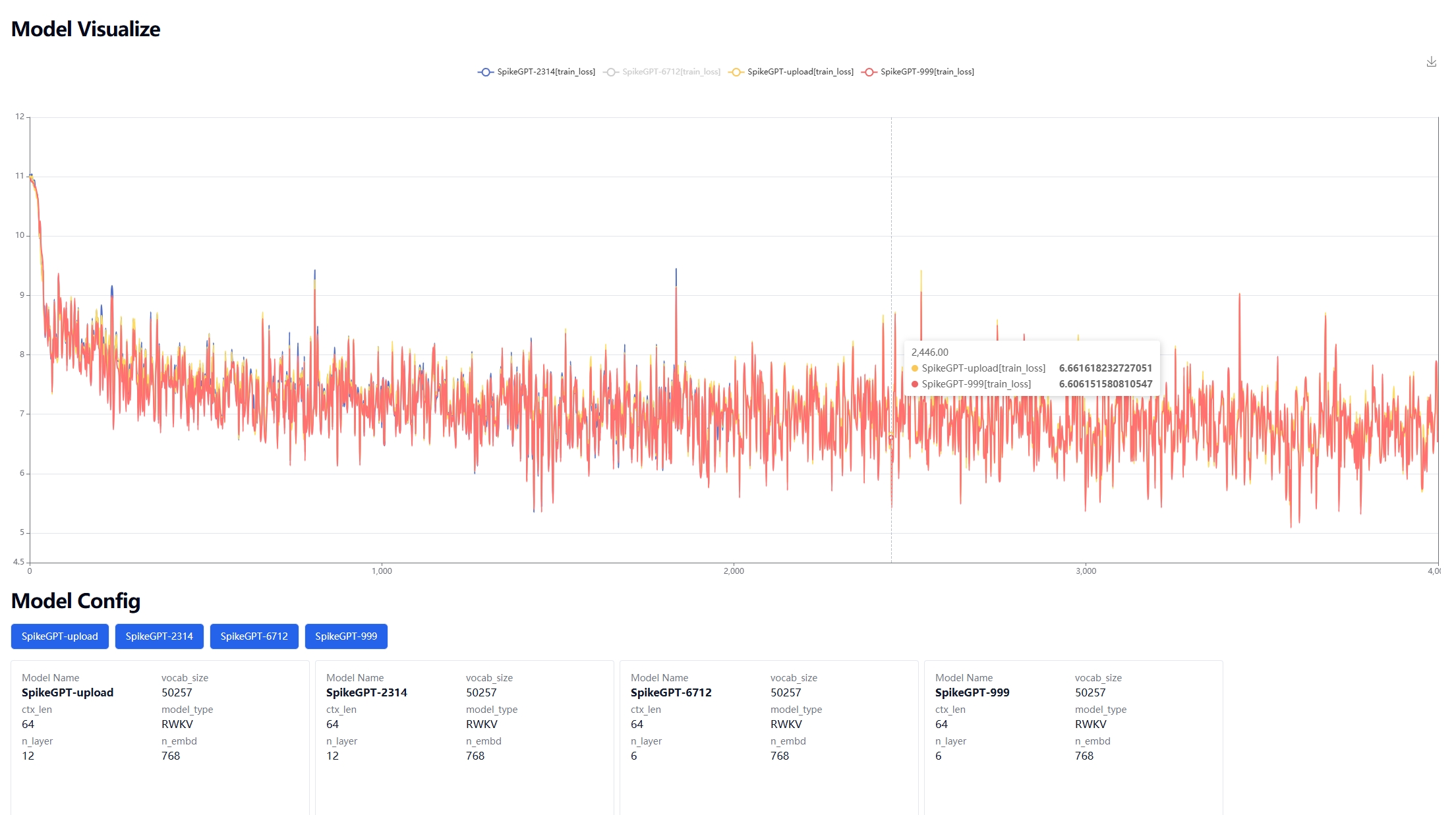} 
        \caption{Visualization of data across multiple models}
        \label{fig:subfigure2}
    \end{subfigure}
    
    \caption{Illustration of logging and visualization tools.} 
    \label{fig:overall_figure} 
\end{figure}

\textbf{10. Unified Interface for Third-Party Extensions: }  Darkit supports seamless integration of third-party contributions. New models, datasets, and modules can be incorporated through a unified plugin system, enabling easy expansion and collaboration (Figure 10).
\begin{figure}[!ht]
    \centering
    \begin{subfigure}[b]{0.48\textwidth} 
        \centering
        \includegraphics[width=\textwidth, height=3.8cm]{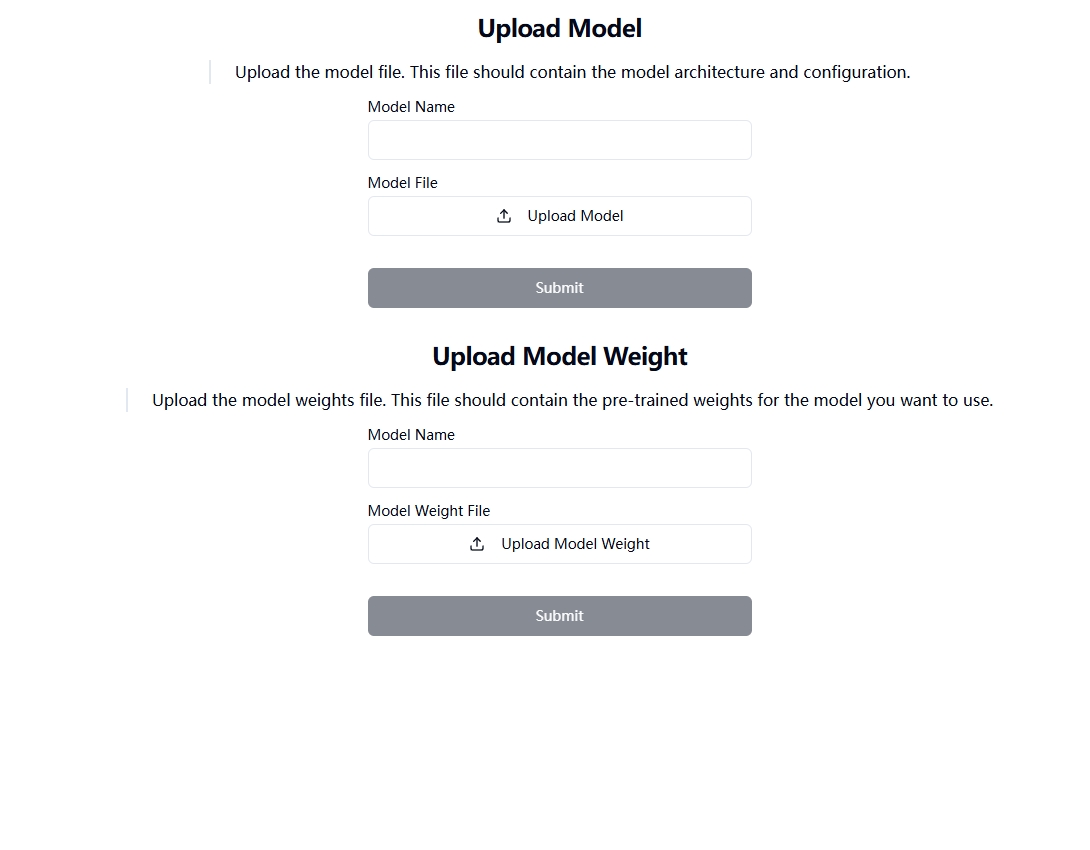} 
        \caption{User interface for uploading models and parameters.} 
        \label{fig:subfigure1} 
    \end{subfigure}
    \hfill 
    \begin{subfigure}[b]{0.48\textwidth}
        \centering
        \includegraphics[width=\textwidth, height=3.8cm]{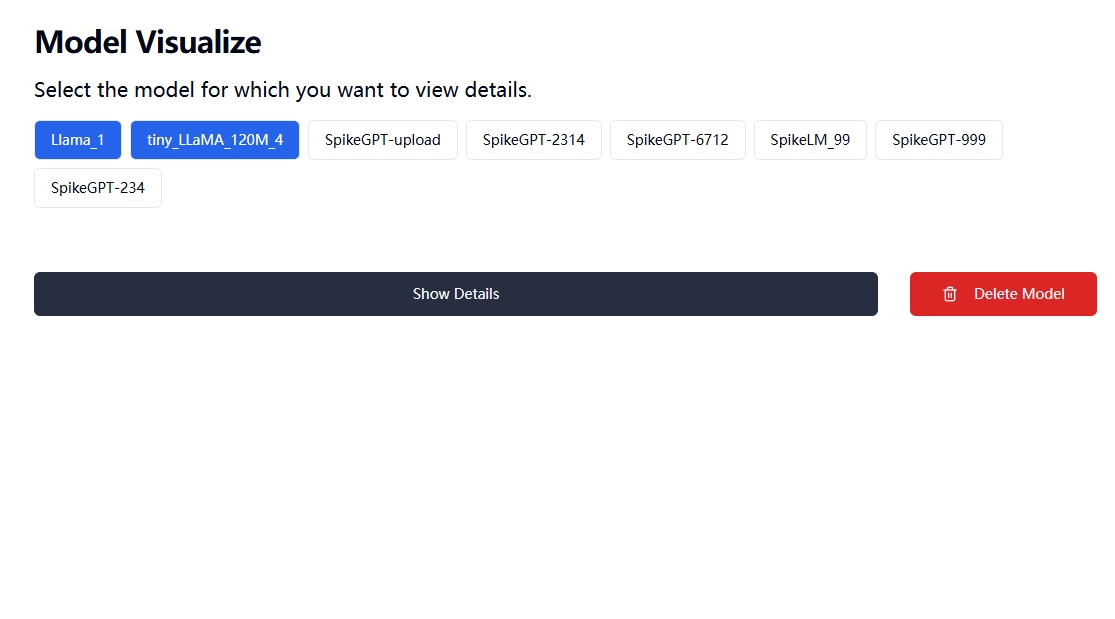} 
        \caption{User-submitted models and parameters ready for execution}
        \label{fig:subfigure2}
    \end{subfigure}
    
    \caption{Workflow for uploading and running custom models and parameters.} 
    \label{fig:overall_figure} 
\end{figure}

Darkit encapsulates and integrates mainstream large language model architectures (such as GPT~\cite{zong2022survey}, BERT~\cite{jawahar2019does}, and Llama~\cite{touvron2023llama}) through standardized application programming interfaces (APIs), providing a flexible, open, and efficient framework that offers comprehensive support for the learning, research, and development of spiking large language models. Darkit provides an accessible web-based version, which can be accessed via the website \url{http://121.40.226.59:8080/}. The platform's frontend is developed using the Svelte framework~\cite{bhardwaz2023svelte}, while the backend is built on the FastAPI stack~\cite{lubanovic2023fastapi}, utilizing asynchronous non-blocking I/O to ensure high performance and stability. Users can also download the source code for local installation from \url{https://github.com/zju-bmi-lab/DarwinKit}.
%


\bibliographystyle{unsrtnat}


\appendix

\end{document}